\documentclass[twocolumn,showpacs,superscriptaddress]{revtex4}

\usepackage{graphicx}

\newcommand{\beq}{\begin{equation}}
\newcommand{\eeq}{\end{equation}}
\newcommand{\barr}{\begin{eqnarray}}
\newcommand{\earr}{\end{eqnarray}}

\begin{document}

\title{Extended DFT+U+V method with on-site and inter-site electronic 
interactions}

\author{Vivaldo Leiria Campo, Jr.}
\email{vivaldo.leiria@gmail.com}
\affiliation{%
Departamento de F\'{\i}sica, Universidade Federal de S\~ao Carlos, 13590-905, S\~ao Carlos, SP, Brazil}
\author{Matteo Cococcioni}
\affiliation{%
Department of Chemical Engineering and Materials Science, 
University of Minnesota, Minneapolis, Minnesota 55455, USA
}%

\date{\today}

\begin{abstract}
In this article we introduce a generalization of the popular DFT+U method 
based on the extended Hubbard model that includes on-site and 
inter-site electronic interactions.
The novel corrective Hamiltonian is designed to 
study systems for which electrons are not completely localized 
on atomic states (according to the general scheme of Mott localization) 
and hybridization between
orbitals from different sites plays an important role. 
The application of the extended functional to archetypal 
Mott - charge-transfer (NiO) and covalently bonded insulators (Si and GaAs)  
demonstrates its accuracy and versatility and the possibility 
to obtain a unifying and equally accurate description for
a broad range of very diverse systems.
\end{abstract}


\maketitle

\section{Introduction}

Due to its moderate computational cost, the simplicity of its formulation,
and the ability to capture the effects of static correlation,
the DFT+U method \cite{anisimov91,anisimov97} 
(by DFT+U we generically indicate Hubbard-corrected Local 
Density Approximation, 
LDA+U, or Generalized Gradient Approximation, GGA+U)
has become quite popular in recent years to study systems
with strongly localized (typically $d$ or $f$) and thus, correlated,
valence electrons. 
DFT+U consists of a simple corrective Hamiltonian, added to the (approximate) 
DFT energy functional, that is shaped on the Hubbard model 
\cite{hubbardI,hubbardII,hubbardIII,hubbardIV,hubbardV,hubbardVI}. 
In its single-band formulation
the Hubbard Hamiltonian can be written as follows:
\beq
H_{Hub} = t\sum_{\langle i,j \rangle,\sigma}(c^{\dagger}_{i,\sigma}
c_{j,\sigma} + h.c.) + U \sum_i n_{i,\uparrow}n_{i,\downarrow} 
\label{hubm}
\eeq
where $\langle i,j \rangle$ denotes nearest-neighbor sites,
$c^{\dagger}_{i,\sigma}, c_{j,\sigma}$, and $n_{i,\sigma}$ are
electronic creation, annihilation 
and number operators for electrons of spin $\sigma$ on site $i$.
The Hubbard model is normally used to describe the behavior of
systems in the limit of strong electronic localization. In these conditions
the motion of electrons is described by a ``hopping" process 
from one atomic site to its neighbors (first term of Eq.~\ref{hubm})
whose amplitude $t$ is proportional to 
the bandwidth of
the system and thus represents the single-particle term of the total energy. 
The Coulomb repulsion, instead, is only accounted for between electrons on 
the same atom through a term proportional to the product of the
occupation numbers of atomic states on the same site.
The effective strength of the ``on-site" Coulomb repulsion is $U$. 
The explicit account for on-site 
electronic repulsions is ideally suited to study 
Mott insulators.
In fact, in these systems 
the insulating character of the ground state stems from the 
dominance of short-range Coulomb interactions (the energy cost of
double occupancy of the same site) over single-particle terms of the
energy (generally minimized by electronic delocalization) \cite{mott70}
that leads to the localization of electrons on atomic orbitals.

Commonly used approximations to the exact DFT energy functional 
generally describe electron-electron interactions using a mean-field-like
approach (i.e., through functionals of the electronic charge density)
which is not ideally suited to represent the Coulomb repulsion
between localized electrons nor their ``correlated" motion. 
Furthermore, the presence of residual
electronic self-interactions in the approximate DFT functionals, 
enhances the tendency towards delocalization. As a result, 
most approximate functionals fail to capture 
the behavior of systems whose electrons are strongly localized.
In spite of its simplicity, the Hubbard-rooted DFT+U approach
has been very effective in alleviating these well known difficulties
and has been successfully applied to
the study of a large number of quite diverse systems. 
In recent years it has also been used as the starting approximation for
more sophisticated computational methods as, 
for example, the Dynamical Mean-Field Theory 
(DFT+DMFT) \cite{kotliar97,liechtenstein98} 
and, more recently, the GW approximation
\cite{kobayashi08,jiang09}. 

The ``on-site" Hubbard Hamiltonian contains all the essential ingredients 
to capture the physics of Mott localization in strongly correlated systems
and, in fact, a good description of these systems is usually obtained with
DFT+U.
A generalized (``extended") 
formulation of the Hubbard model, with Coulomb interaction terms
between electrons on neighbor sites, has been considered since the
early days of the Hubbard model \cite{hubbardIV,hubbardV}.
However, to the best of our knowledge, it has never been implemented 
in any DFT-based functional, nor used in ab-initio calculations.
A simplified formulation of the original extended Hamiltonian 
(with only charge interactions between couples of sites)
is given in the following expression:
\barr
H_{Hub} &=& t\sum_{\langle i,j \rangle,\sigma}(c^{\dagger}_{i,\sigma}
c_{j,\sigma} + h.c.) + U \sum_i n_{i,\uparrow}n_{i,\downarrow} \nonumber \\ 
&+& V \sum_{\langle i,j \rangle}n_{i}n_{j}
\label{hubvm}
\earr
where $V$ represents the strength of the interaction between
electrons on neighbor sites $i$ and $j$.

The use of the extended model has been stimulated in the last decades
by the discovery of high T$_c$ superconductors 
and the intense research activity focusing
around them.
Whether the inter-site coupling $V$ has a determinant role
in inducing superconductivity is still matter of debate.
The ``resonating valence bond" model \cite{andersonsci87}
predicts a superconducting state (at least within
mean-field theory) for a doped Mott insulator with only on-site
couplings \cite{andersonprl87}. However, several numerical studies 
suggest that the inter-site interaction plays indeed an important role
\cite{imada91,hirsch88} and superconductivity is predicted
in a regime with repulsive on-site ($U > 0$) and attractive 
inter-site ($V < 0$)
couplings \cite{thakur07,jursa96,szabo96,demello99}.
Several studies have also demonstrated that
in the ``normal" (non-superconducting) state of superconductors
and, in general, in correlated materials, the relative strength
of $U$ and $V$ controls many properties of the ground state
as, for example, the occurrence of possible phase separations \cite{mancini09},
the magnetic order \cite{morohoshi08,watanabe08}, 
the onset of charge-density and spin-density wave regimes \cite{vandogen94}.
More recently, the extended Hubbard model has been used to study the conduction
and the structural properties of polymers and carbon 
nano-structures and the interplay between $U$ and $V$ was shown
to control, for example, 
the dimerization of graphene nanoribbons \cite{zhu06}.
The importance of a more accurate account of inter-site couplings
and of the detailed balance with on-site interactions
has been also recognized in a theoretical study of Fe impurities
deposited on various alkali metal films \cite{carbone09}.
In this case, however, the hybridization between the impurity $d$ states
and the $s$ and $p$ orbitals of the alkali metal substrate
was described using a one-body term in the Hamiltonian (effectively
corresponding to a generalized hopping process between the localized
and the delocalized states) rather than a two-body inter-site 
electronic interaction.

In this work we propose to introduce the extended
Hubbard model in DFT-based calculations through 
a generalization of the corrective functional DFT+U is based on
to include on-site and inter-site electronic interactions.
The main aim of the novel formulation is to improve the accuracy
of the DFT+U scheme and to make it quantitatively predictive 
for systems where correlation is not strong enough to induce a
complete Mott localization of electrons on atomic states, or for which,
in general, the hybridization between orbitals belonging to different atoms
plays an important role in determining the properties of the ground state.
The novel functional is tested on quite diverse bulk solids as NiO, Si and
GaAs.
In NiO, a prototype strongly correlated material, 
$d$ states are quite well localized around Ni atoms.
Nevertheless their hybridization with O $p$ states
plays a quite significant role and is 
one of the main factors to determine the 
charge-transfer insulating character of the material.
Si and GaAs are, on the opposite extreme, typical band insulators.
The hybridization between $s$ and $p$ orbitals belonging to the same
and to neighbor sites plays a dominant role in determining the 
electronic structure of these materials as it leads to 
the formation of bonds in between atoms (with tetrahedral symmetry) and to the 
consequent onset of the semiconducting character of these systems
due to the splitting between bonding (valence) and
anti-bonding (conduction) states.
The choice of these systems is not casual:
a computational scheme that is able to describe accurately strongly localized
as well as ``strongly hybridized" systems is likely to be successful for 
intermediate situations as well that are, by far, the most difficult 
to deal with.
For each of these test systems we study electronic and structural properties
comparing the results from the new functional with those obtained
from ``standard" DFT (GGA) and DFT+U calculations. Improvements obtained 
with the novel formulation and still remaining issues will be highlighted in 
each case.

The paper is organized as follows: in section II the extended DFT+U+V
energy functional is introduced and discussed.
In section III we present the results obtained from the application
of this novel approach to the study of bulk NiO, Si and GaAs.
Finally, some conclusive remarks are proposed.

\section{The extended DFT+U+V functional}
The LDA+U approach was formulated and developed by Anisimov and 
coworkers in the first part of the 1990s \cite{anisimov91,anisimov97}
to improve the accuracy of LDA in describing systems characterized
by localized, strongly correlated valence electrons.
This scheme is based on a correction to the DFT energy functional that 
can be generally written as follows:
\beq
E_{DFT+U} = E_{DFT}+E_{U} = E_{DFT}+E_{Hub}-E_{dc}.
\label{eu}
\eeq
In this equation E$_{Hub}$ is the part that
contains electron-electron interactions 
as modeled in the Hubbard Hamiltonian.
$E_{dc}$ is a mean-field approximation to
$E_{Hub}$ and models the amount of electronic correlation
already contained in E$_{DFT}$.
This term is subtracted from the total functional to avoid
double-counting of the electronic interactions contained in 
$E_{Hub}$.

Following a commonly used simplified form \cite{dudarev98} of the rotationally
invariant formulation of DFT+U introduced by Liechtenstein 
et al., \cite{liechtenstein95} the total corrective functional results:
\barr
\label{ehub}
E_U&=&E_{Hub} - E_{dc}
= \sum_{I,\sigma} \frac{U^I}{2}Tr\left[{\rm \bf n}^{I\sigma}
\left({\bf 1}-{\rm \bf n}^{I\sigma}\right)\right]. \
\earr
In this equation ${\rm \bf n}^{I,\sigma}$ is the ``on-site" occupation matrix 
that is defined by the projection of the occupied Kohn-Sham orbitals
of spin $\sigma$ on the localized (atomic) states of orbital quantum number 
$l$ (typically of $d$ or $f$ kind) $\phi^I_m$ :
\beq
\label{occup}
n^{I\sigma}_{m,m'}=\sum_{k,v}f_{kv}\langle \psi_{kv}^{\sigma}|
\phi_{m'}^I\rangle\langle \phi_m^I | \psi_{kv}^{\sigma} \rangle.
\eeq
In Eq.~(\ref{occup}) $m$ is the magnetic quantum number associated with
$l$ ($-l\leq m\leq l$), while $f_{kv}$ are the occupations
of the Kohn-Sham orbitals $\psi_{kv}^{\sigma}$ based on the
distribution of their energies around the Fermi level.

The definition of the novel DFT+U+V functional can be 
obtained from a formal generalization of the atomic orbital
occupation matrix in Eq.~(\ref{occup}), to allow for the possibility that the
two atomic wavefunctions involved in the definition belong to different atoms:
\beq
\label{occupij}
n^{IJ\sigma}_{m,m'}=\sum_{k,v}f_{kv}\langle \psi_{kv}^{\sigma} | \phi_{m'}^{J}  
\rangle\langle \phi_m^{I} | \psi_{kv}^{\sigma}\rangle.
\eeq

In Eq.~(\ref{occupij}) 
the indexes $m$ and $m'$ run over the angular momentum manifolds 
that are subjected to the Hubbard 
correction on atoms $I$ and $J$ respectively. 
It is important to notice that the 
occupation matrix defined in Eq.~(\ref{occupij})
contains informations about all the atoms
in the same unit cell and the on-site occupations defined 
in Eq.~(\ref{occup}) correspond to its diagonal blocks.

Using the occupation matrix of Eq.~(\ref{occupij})
the energy functional in Eq.~(\ref{ehub})
can be generalized as follows:

\barr
\label{UV}
E_{UV}=E_{Hub} - E_{dc} &=& \sum_{I,\sigma} \frac{U^I}{2}Tr\left[{\rm \bf n}^{II\sigma}
\left({\bf 1}-{\rm \bf n}^{II\sigma}\right)\right]\nonumber \\
&-& \sum_{I,J,\sigma}^{\hspace{5mm}*}\frac{V^{IJ}}{2} 
Tr\left[{\rm \bf n}^{IJ\sigma}{\rm \bf n}^{JI\sigma}\right]
\earr

where, specifically,

\beq
\label{dcv}
E_{dc} = \sum_I \frac{U^I}{2}n^I(n^I-1) +
\sum_{I,J} \frac{V^{IJ}}{2} n^{II}n^{JJ}
\eeq
\barr
\label{hubv}
E_{Hub} &=& \sum_I \frac{U^I}{2} \left[\: (n^I)^2 -
\sum_\sigma Tr\left[({\rm \bf n}^{I\sigma})^2\right] \:\right] \nonumber \\
&+&\sum_{IJ} \frac{V^{IJ}}{2} \left[n^{II}n^{JJ} -
\sum_\sigma Tr({\rm \bf n}^{IJ\sigma} {\rm \bf n}^{JI\sigma})\right].
\earr

In Eq.~(\ref{UV}), $U^I$ and $V^{IJ}$ represent on-site and inter-site 
interaction parameters respectively. The star in the second sum operator 
reminds that for each atom $I$, index $J$ covers all 
its neighbors up to a given distance (or belonging to the same shell)
and the trace operator $Tr$ is understood as follows: 
$Tr\left({\rm \bf n}^{IJ\sigma}{\rm \bf n}^{JI\sigma}\right) = 
\sum_{m,m'} n^{IJ\sigma}_{mm'}n^{JI\sigma}_{m'm}$. 
In Eqs.~(\ref{dcv}) and (\ref{hubv}), $n^I = n^{II} = \sum_\sigma
Tr\:\left[{\rm \bf n}^{II\sigma}\:\right]$ is the total number of 
electrons on the atomic states of site $I$.

Having been obtained as the extension of a simplified 
model (Eq.~(\ref{ehub})), the inter-site term of the DFT+U+V corrective
functional (Eq.~(\ref{UV})) necessarily 
corresponds to a quite drastic approximation 
of the full off-site many-body part of the electronic Hamiltonian.
In fact, terms involving states belonging to three or four distinct atoms
are all neglected. In addition, among the two-site terms 
only a restricted subset is explicitly included in the extended 
corrective Hamiltonian. If we establish the notation:
$A^{IJKL}_{ijkl}=\langle \phi^I_i\phi^J_j|V_{ee}|\phi^K_k \phi^L_l\rangle
\left(n^{IK,\sigma}_{ik}n^{JL,\sigma'}_{jl}-\delta_{\sigma\sigma'}
n^{IL,\sigma}_{il}n^{JK,\sigma'}_{jk}\right)$, where $V_{ee}$ corresponds
to the kernel of the electronic interaction, and $\phi^I_i$ represents
the state $i$ of atom $I$, the Coulomb part of the term 
included in our inter-site functional
corresponds to $A^{IJIJ}_{ijij}$ (averaged over the orbitals of the two sites). 
Other contributions
describing cross charge exchanges between neighbor sites ($A^{IJJI}_{ijji}$), 
double (parallel) electron transfers from one site to a neighbor one 
($A^{IIJJ}_{iijj}$), couplings between on-site charge and hopping from/to
the same site $A^{IIIJ}_{ii'ij'}$, 
and coupled intra-site charge exchanges
(as, e.g., $A^{IJIJ}_{iji'j'}$) are all neglected. The one included in our
corrective functional (proportional the product of number operators 
on different sites) is thus the only one that has a classical Coulomb 
counterpart and is, presumably, the most 
relevant one. This ansatz, not explicitly tested in this work,
is indeed consistent with the approximation used in the 
formulation of the model \cite{hubbardIV,hubbardV} and in the very 
abundant literature where the extended Hubbard model is used. 

As evident from Eq.~(\ref{ehub}), the
on-site term of the total energy introduces a finite energy cost 
for fractional atomic 
occupations (if $\lambda^{I,\sigma}_m $ is one eigenvalue of $n^{II\sigma}$, 
$E_U$ is 0 for $\lambda^I_m = 0$ or $\lambda^I_m = 1$, positive otherwise); 
this penalty favors a Mott-like ground state with correlated 
electrons localized on atomic orbitals.
The effect of the inter-site interaction can be easily
understood from the contribution of the corrective
functional in Eq.~(\ref{UV}) to the total Kohn-Sham potential.
This quantity can be computed as the functional derivative 
of the energy with respect to $(\psi_{kv}^{\sigma})^*$:
\barr
\label{VUV}
&&V_{UV}|\psi_{kv}^{\sigma}\rangle  = 
\frac{\delta E_{Hub}}{\delta (\psi^{\sigma}_{kv})^*}\nonumber \\
&&= \sum_I \frac{U^I}{2}\sum_{m,m'}\left(\delta_{mm'} -
2{\rm \bf n}^{II\sigma}_{m'm}\right)|\phi^{I}_{m}\rangle\langle\phi^{I}_{m'}|\psi_{kv}^{\sigma}\rangle \nonumber \\
&&- \sum_{I,J} V^{IJ}
\sum_{m,m'}{\rm \bf n}^{JI\sigma}_{m',m}|\phi^{I}_{m}\rangle\langle\phi^{J}_{m'}|\psi_{kv}^{\sigma}\rangle.
\earr
From Eq.~(\ref{VUV}) it is evident that the {\it on-site}
term of the potential is attractive for occupied states that are, at most,
linear combinations of atomic orbitals of the {\it same atom}
(resulting in on-site blocks of the occupation matrix, 
${\rm \bf n}^{II\sigma}$, dominant on others),
whereas the {\it inter-site} one stabilizes states that are linear
combinations of atomic orbitals belonging to {\it different atoms} 
(e.g., molecular orbitals, that lead to large off-site blocks, 
${\rm \bf n}^{JI\sigma}$, of the occupation matrix). 
Thus, a competition sets between two 
opposite tendencies that allows for more general localization
regimes and increases the coupling between orbitals on different 
sites. Obviously, the character of the electronic ground state
depends critically on the relative strength of the on-site ($U$) 
and the inter-site ($V$) electronic 
interactions. The detailed balance between these quantities is guaranteed 
by the possibility of computing both parameters simultaneously
through the linear-response approach described in \cite{cococcioni05}. 
In fact, the inter-site interaction parameters correspond to the
off-diagonal terms of the interaction matrix defined in Eq.~19
of Ref.~\cite{cococcioni05}.

It is important to notice that
the trace operator in the on-site functional guarantees 
the invariance of the energy under rotations of atomic orbitals 
{\it on the same atomic site}. However, in the inter-site term, the trace
applied to the product of generalized occupation matrices, including orbitals
from multiple sites, is not sufficient to strictly enforce the 
invariance with respect to rotations 
involving all atomic orbitals in the system at the same time 
(i.e., mixing atomic orbitals
that are centered on {\it different sites}). In fact, to obtain this
covariance, the elements of the electronic interaction matrix should transform
as quadruplets of atomic orbitals and thus have full
site- and orbital- dependence.
At the present level of approximation, 
interactions between more than two atomic sites 
are assumed to be less important than the ones included 
in Eq.~(\ref{UV}) and orbital dependence is totally neglected
(the effective interactions included in the present formulation
can be seen as atomic averages of the fully orbital-dependent quantities). 
This assumption, consistent with the one operated in the original formulation 
of the extended Hubbard model \cite{hubbardIV,hubbardV},
appears reasonable for the basis set of atomic orbitals, 
but may not be generally valid if a different
basis is used to represent localized states.
The inclusion of inter-site couplings is, however, a significant 
step towards general covariance with respect to the on-site functional.
Furthermore, the possibility
of evaluating both $U$ and $V$ from linear-response theory \cite{cococcioni05} 
(at least for interactions between couples of atomic sites), 
guarantees a high level of consistency between the atomic orbital
basis set and the interaction parameters used in the functional, and
reduces the dependence of the results on the specific
choice of the localized basis.
Site- and orbital-dependence of the corrective functional is implicitly
included in Wannier-function-based implementations of the on-site DFT+U
\cite{mazurenko07} if the effective interactions are computed 
consistently. However, even in a (e.g., maximally localized)
Wannier-function-based formulation, no guarantee exists that inter-site
or multiple-site terms of the corrective functional can be neglected
(in fact the Wannier basis is not generally constrained to be the one 
on which the interaction is site-diagonal).
With respect to this approach, a formulation based
on atomic states has the advantage of not requiring the preliminary 
calculation of the localized basis orbitals and offers a physically
transparent way to select a minimal number of relevant 
interaction parameters to be computed (e.g., based on distances between 
neighbors).

In the implementation of Eq.~(\ref{UV}) we have added 
the possibility for the corrective functional to act
on two $l$ manifolds per atom as, for example,
orbitals $3s$ and $3p$ in Si, or orbitals $4s$ and $3d$ in Ni.
To the best of our knowledge, this feature has been implemented only
recently in a ``standard" (on-site) DFT+U functional 
\cite{paudel08} where, however, at variance with our formulation,
no interaction was established between the two manifolds of the same atom.
If we call ``standard" the higher $l$ states of each atom the Hubbard 
correction acts on, and ``background" the other $l$
manifold being corrected, for a given couple of atomic sites $I$ and $J$ 
(Eq.~(\ref{UV})), we have four interaction parameters: 
standard-standard, standard-background, background-standard, 
and background-background. 
Alternatively, the different Hubbard-corrected manifolds on each atom can 
be seen as belonging to different atoms located at the same 
crystallographic site. 
The motivation for this extension consists in the fact that 
different manifolds of atomic states may require 
to be treated on the same theoretical ground in cases where hybridization 
is relevant (as, e.g., for bulk diamond whose bonding structure 
is based on the $sp^3$ mixing of $s$ and $p$ orbitals). 
The particular choice to have different interaction parameters
acting on different manifolds of atomic states, and the possibility
to include ``cross-manifold" interactions (basically ``on-site" $V$ parameters 
between standard and background states) eliminates the need for
``ad-hoc" constructions of the basis set 
for cases where localization is expected to occur on states different 
from atomic orbitals (see, for example, \cite{kresseCO03},
\cite{kresseCO04}, \cite{dabo07}).

While the inter-site interaction $V$ can be calculated at the same time
of the on-site $U$ and with no additional cost,
the computational workload of a DFT+U+V calculation depends on the
number of neighbors between which the inter-site correction is established.
For interactions between nearest neighbors (all contained in a 
$3\times3\times3$ supercell centered around the primitive one) 
the computational cost is only marginally bigger than for an 
on-site-only DFT+U; if further shells of neighbors
are included in the summation of Eq.~\ref{UV}, 
larger supercells around the primitive one may be needed and the search
of equivalent neighbors may increase significantly the cost
of the calculation (up to a 18\% increase in the 
cpu time for a $5\times5\times5$ supercell
compared to a $3\times3\times3$).

\section{Results and Discussions}
In this section we present the results obtained with the novel functional 
applied to the study of the selected test systems: NiO, Si, and GaAs.
While the choice to test a numerical approach explicitly designed
for strongly correlated materials on band insulators could appear unusual,
it is important to notice that, for both classes of materials, the fundamental
gap has largely the same nature: it is mostly due
to the derivative discontinuity of the exchange-correlation 
energy functional (see, for example, \cite{perdew83,grosslibro}) that
is missing from standard DFT approximations.
Thus a correction, as DFT+U, ideally designed to introduce this feature in the 
energy spectrum of a system, should have general validity for semiconductors
and insulators in both families. 
Analogous considerations have led the authors of \cite{grossprb08} 
to operate a similar selection of systems to test their Reduced 
Density Matrix Functional (RDMF) approach.

The use of the novel functional to study NiO is motivated by the
need to validate it on systems (as Mott or charge-transfer insulators) 
for which the on-site DFT+U already gives reasonably good results.
Also, the charge-transfer character of the band gap of this material
(i.e., the hole in the O ligand field observed in photoemission 
experiments) \cite{sawatzky84} suggest a more significant hybridization
between $d$ and $p$ states than in other oxides.
Si and GaAs are, instead, typical band insulators, with covalent
and ionic character, respectively. 
The application to these materials is a very important one for the
novel functional introduced in this work as 
the electronic structure of these systems is based on
the formation of bonds and thus on the hybridization
between states of neighbor atoms.

For all these systems comparison will be made with the results obtained with
DFT and with the ``standard", on-site DFT+U. 
All the calculations were performed using the 
plane-waves pseudo-potential ``pwscf" code contained in the 
$Quantum~ESPRESSO$ (QE) package \cite{qe,qe1}, where we 
implemented the ``+U+V'' correction starting from an existing
on-site DFT+U functional \cite{qe1}. Except in one case (detailed
in the paper), all the pseudopotentials were obtained from the QE 
pseudopotentials database \cite{qe}.

It is important to remark that for all the systems treated in this work
$U$ and $V$ were obtained using the linear response approach described in
\cite{cococcioni05}. 
However, at variance with the calculations presented in \cite{cococcioni05}, 
the response matrices were not constrained to give a neutral
total response (in terms of on-site occupations) when the
perturbation is applied on every atom. 
While this has no effects on the final results
if big enough supercells are used in the calculations, we 
believe it to be a better strategy to compute
the effective interactions especially if the response 
of all atomic orbitals (standard plus background) is
explicitly considered. 
A further difference with the results presented in \cite{cococcioni05} consists
in the use of orthogonalized atomic orbitals to construct occupation matrices.
While not necessary, this choice guarantees 
that the atomic occupation matrices satisfy more stringent sum rules
(their trace is closer to the
total number of electrons on the atomic states). Some small differences
are to be expected with the results obtained for NiO in \cite{cococcioni05}; 
however, we
believe that the consistent evaluation of the effective electronic
interactions (especially if obtained from a DFT+U ground state) 
reduces these differences to a minimal value.

\subsection{NiO}

As other transition-metal oxides with the same stoichiometry, 
nickel oxide has a cubic rock-salt crystalline structure. 
However, below a Ne\'el temperature of 523 K, the magnetic moments of Ni
atoms arrange in an antiferromagnetic (AF) 
order (usually referred to as AFII) where
ferromagnetic (111) Ni planes alternate with opposite magnetization.
As a consequence of the AF magnetic order the crystal acquires a 
rhombohedral symmetry. 
Transition metal oxides (TMOs) have represented 
a significant challenge for theorists since their insulating character can
not be explained satisfactorily using band theory due to the intrinsic
many-body origin of their band gap.
Although within Kohn-Sham theory the energy spectrum has no precise physical
meaning and is not bound to reproduce the band gap of the system,
once the finite discontinuity of the (exact) exchange-correlation
potential is added to the HOMO-LUMO energy 
difference the correct band should be exactly reproduced 
\cite{perdew83,grosslibro}. 
Failing to properly incorporate many-body effects, most of approximate
exchange-correlation functionals produce no discontinuity in the
corresponding potentials, thus resulting in quite poor estimates of
band gaps or in their total suppression.
In fact, many TMOs are predicted to be metal in striking contrast 
with the observed insulating character. In some of these systems
(e.g., FeO \cite{cococcioni05}) lack of proper account
of many-body aspects of the electronic structure also results in
serious inaccuracies in the description of structural and magnetic
properties.
Similar problems can be expected for NiO even if a gap 
(still significantly smaller than the experimental one) fortuitously
opens in its Kohn-Sham spectrum as the result of the 
balance between the exchange and the crystal-field splittings 
among the $d$ states of Ni (nominally occupied by eight electrons). 
Photoemission experiments on NiO have measured a band gap of 
about 4.3 eV (3.1 eV at the minimum intensity) \cite{sawatzky84}
and have explored several features of the spectrum assessing,
in particular, its charge-transfer character \cite{sawatzky84,thuler83,
fujimori84}. In fact,
the excitation of one electron across the band gap of the quasi-particle
spectrum corresponds to its transfer from the $p$ states of a ligand oxygen
atom to the $d$ states of a Ni atom. As a consequence, the top
of the valence band is dominated by the $p$ states of oxygen while the bottom
of the conduction one largely consists of Ni $d$ states.
Computational studies on this material have been quite successful in 
reproducing these features of its spectrum.
DFT+U, whose corrective potential is designed to introduce a finite 
energy difference
between occupied and unoccupied states, can be expected
to produce a more accurate estimate of the fundamental gap of a system
(in principle larger than the one appearing in photoemission
experiments). In fact, it has been used quite successfully to study
NiO and has produced a band gap of about 3.0 to 3.5 eV (the precise value
varies with the $U$ used in different works) and quite 
accurate estimates for both the magnetic moments and the equilibrium
lattice parameter \cite{dudarevnio00,bengone00,kressenio04}.
For other details of the density of states (as, e.g., the spectral 
weight of O $p$ states on top of the valence band), 
the agreement was not unanimous.
DFT+U has also been employed recently to compute the $k$-edge 
XAS spectrum of NiO 
using a novel, parameter-free computational approach \cite{maurinio09}
that has produced results consistent with experimental data.
In this articles authors highlight 
the importance of non-local excitation of $d$ states
and, specifically, those
involving second-nearest neighbor Ni atoms.
The same intent of improving the description of spectroscopic properties 
has been pursued with the GW approximation based on
a DFT+U ground state \cite{kobayashi08,jiang09}; 
this approach has provided a better estimate
of the energy gap compared to DFT+U even though 
other details of the density of states were almost unchanged 
\cite{kobayashi08}. 
Hartree-Fock (HF) \cite{martin02} and hybrid functional (e.g. B3LYP)
\cite{martin02,feng04} have also been used to study NiO. While pure HF
overestimates the size of the band gap and fails to reproduce
its charge-transfer nature, hybrid functionals are more accurate
and result in a better estimate of the properties of this material.
Most recently DFT+DMFT calculations of NiO have explored the
effects of dynamical correlations and have produced a band gap
in the energy spectrum in excellent agreement with photoemission
experiments, even though its charge-transfer character was not always 
well reproduced \cite{vollhardt06,kunes07}.
A modification to this method within the iterative perturbation
theory has been recently used to fix this particular aspect of the 
calculated spectrum of NiO \cite{miura08}. In this paper, the authors
also present a detailed discussion of the features of the DOS 
remarking the role of dynamical correlations (e.g., in eliminating
the excessive spectral weight of mixed Ni $e_g$ and O $p$ states
at the bottom of the valence band, normally observed in DFT+U and GW),
and the importance that inter-spin and inter-atomic components of
the self-energy would have in fixing some still remaining discrepancies 
in the computed spectrum compared to the experimental one (e.g.,
the position of the occupied satellite). 
This latter hypothesis will be discussed with the results of the present work
where inter-atomic couplings are explicitly taken into account thanks to the
extended functional introduced in the previous section.

All calculations presented in this paper were performed in the AF 
configuration of the system.
We used ultrasoft \cite{vanderbilt90} GGA pseudopotentials constructed with
the PBE parametrization \cite{pbe}
(Ni.pbe-nd-rrkjus.UPF and O.pbe-rrkjus.UPF of the
Quantum-Espresso repository \cite{qe}) and based on the following
valence configurations: $3d^84s^24p^0$ for Ni and $2s^22p^4$ for O. 
Energy cut-offs of 40 and 400 Ry were used
for wavefunctions and charge density respectively.
A $4\times4\times4$ Monkhorst-Pack $k$-point grid \cite{monkhorst76} 
was used to sample
the Brillouin zone of this system. 
Due to its magnetization, NiO was formally treated as a metal with a 
0.01 Ry gaussian smearing of its Fermi distribution function that 
conveniently placed the (fake) Fermi level in 
the middle of the energy band gap.

\begin{figure}
\vspace{-0.1in}
\includegraphics[width=5.5cm,keepaspectratio,angle=-90]{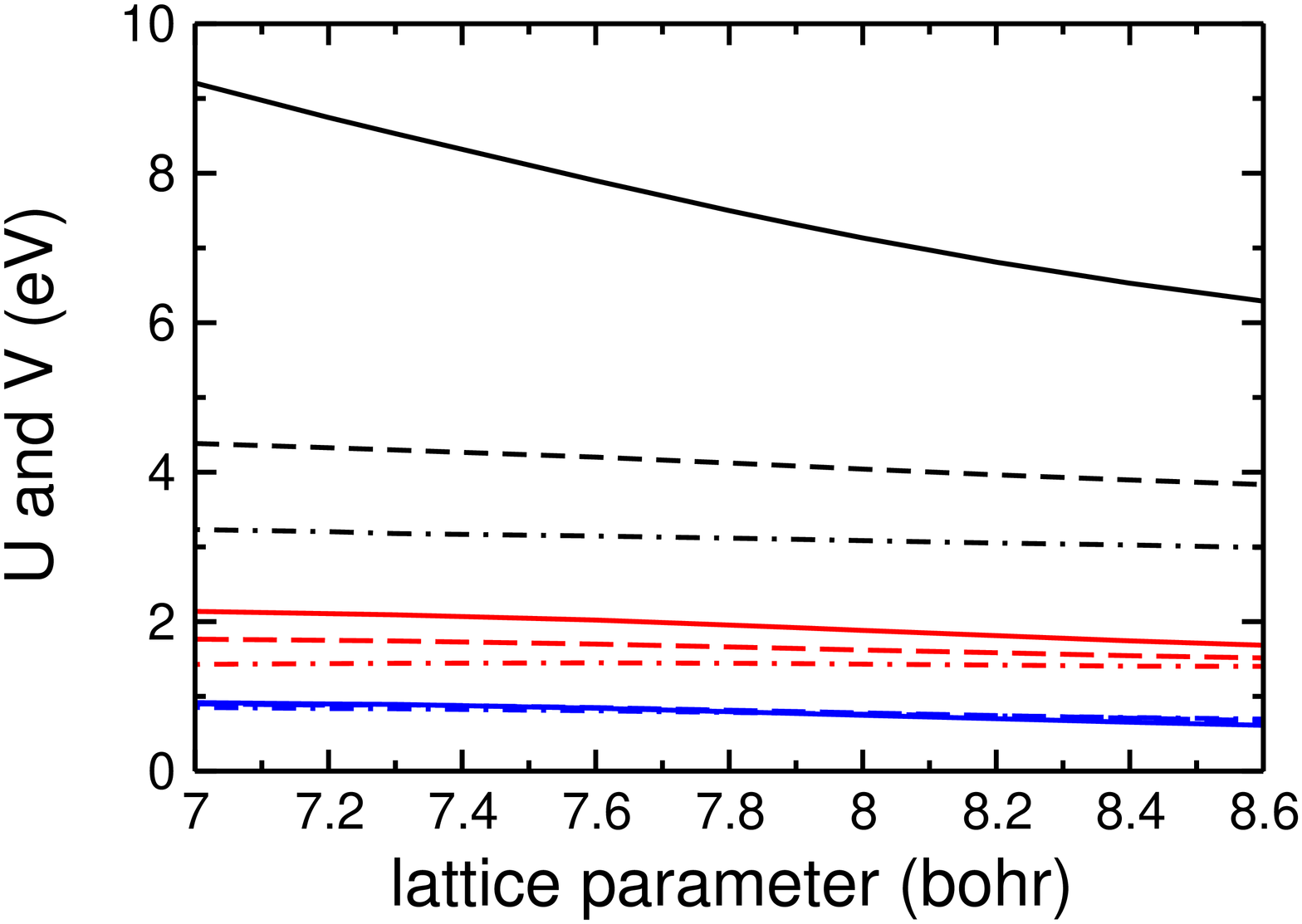}
\vspace{-0.1in}
\includegraphics[width=5.5cm,keepaspectratio,angle=-90]{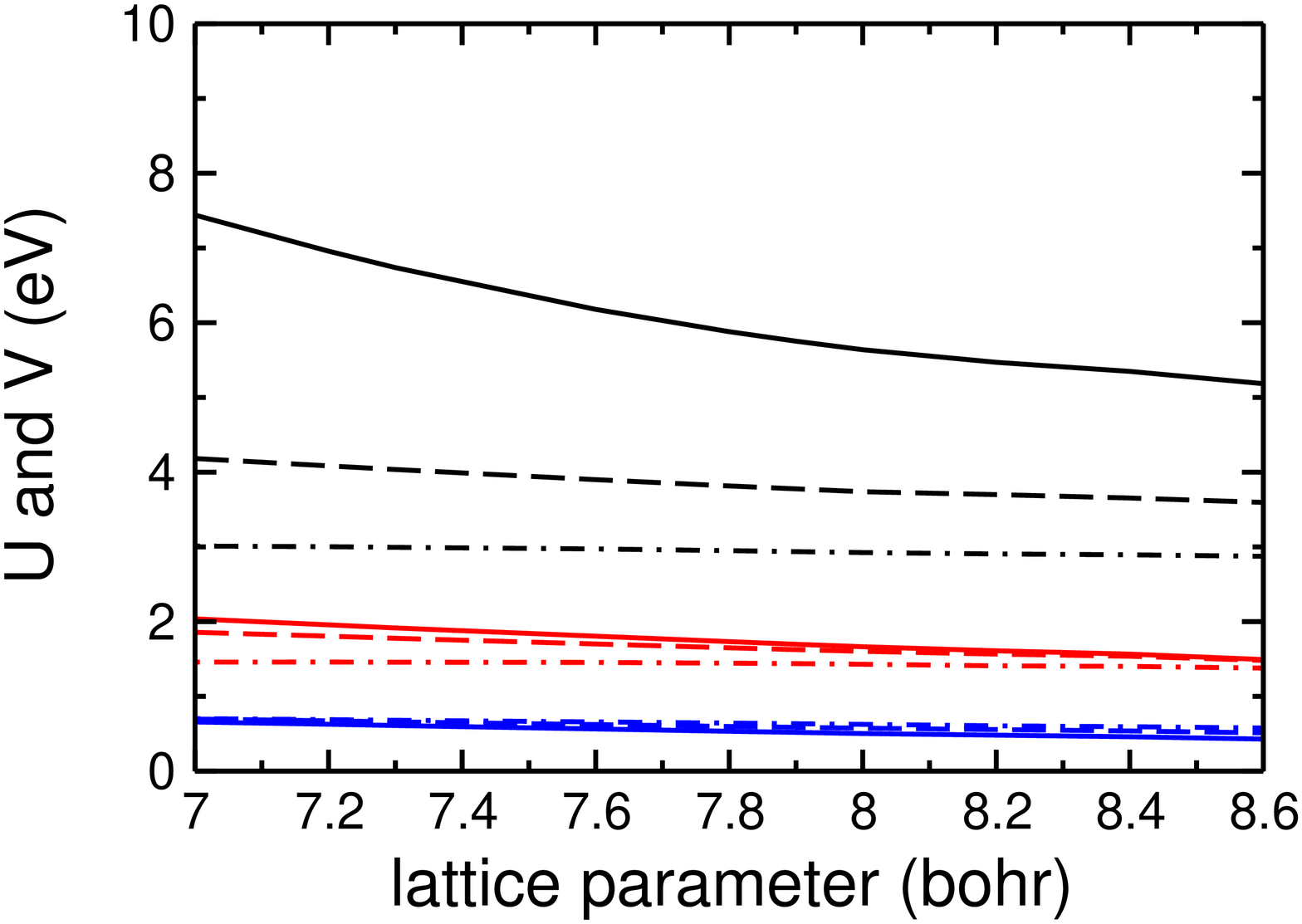}
\caption{\label{uva0} (Color online) 
The variation of the on-site and inter-site electronic couplings
as functions of the cubic lattice parameter of NiO. 
The interaction parameters were obtained from a GGA ground state
(top graph) or from a GGA+U+V ground state (bottom graph) according
to a self-consistent procedure (see text). Black lines represent 
on-site (Ni only) interactions, red ones describe the couplings 
between nearest neighbors (Ni-O), blue ones between next-nearest 
neighbors (Ni-Ni). Solid lines are for standard-standard interactions,
dashed for standard-background, dot-dashed for background-background.
}  
\end{figure}

The electronic interaction parameters ($U$ and $V$) used in this work 
were re-computed for each considered lattice spacing (using
the linear-response approach of ref. \cite{cococcioni05}). Fig.~\ref{uva0} 
shows the dependence of some of these parameters
on the size of the unit cell. 
It is important to remark that 
these parameters are not directly comparable with those 
evaluated in \cite{cococcioni05} 
for two reasons: first, in the present calculations
we have explicitly included the response of the ``background"
states, which could not be done in the previous work; second, 
at variance with what was done in Ref. \cite{cococcioni05} to define
the on-site occupation matrix we have used atomic orbitals that
were preliminarily orthogonalized. 
However, since $U$s and $V$s are consistently computed and used within 
the same approximation (i.e., using the same definition of the occupations)
in both works the most significant differences 
are expected to descend from using a
$V$-augmented functional.

As it can be observed from Fig.~\ref{uva0} (top graph) the on-site $U$ 
is the parameter that changes the most with the size of the unit cell.
Other electronic couplings are much less affected and the dependence
of the interaction between second-nearest neighbors (Ni-Ni) on the lattice
parameter is almost negligible.
While the parameters shown in the top graph of figure \ref{uva0} were 
obtained from the GGA ground state of the system, those represented in
the bottom graph were computed ``self-consistently" from a GGA+U+V ground
state using a method equivalent to that of ref. \cite{kulik06} but
based on a more efficient algorithm that will be described elsewhere.
Although all the electronic interactions are obtained
at the same time (including those for ``background" states and their coupling
with ``standard" ones) the interaction parameters between the $p$ states
and between $p$ and $s$ states of oxygen (all $U_O$ and $V_{O-O}$) 
were excluded from our DFT+U+V calculations. 
While a rigorous motivation for this choice is presently missing,
we believe that the corrective action of the ``+U+V" functional
is mostly needed for open manifolds around the Fermi level (or
the top of the valence band) while $p$ and $s$ states of oxygen are almost
completely full.
Besides the on-site $U_{Ni}$, first ($V_{Ni-O}$) and second
($V_{Ni-Ni}$) nearest neighbor inter-site couplings (between 
``standard" and ``background" states) were 
all included in our calculations; the inclusion of the interactions between
further neighbors was found to produce no relevant effect. 
For all the approaches compared in this work (GGA, GGA+U and GGA+U+V)  
the optimization of the rock-salt cubic structure (in the AF 
ground state) was performed using the cell-size dependent electronic
couplings plotted in Fig.~\ref{uva0}, fitting the the dependence of the 
total energy on its volume on a Murnaghan equation of state. 
Using size-dependent electronic couplings has been recently shown to be
fundamental for quantitative descriptions of structural and electronic 
properties of materials and for accurate evaluations of transformation
pressures between, e.g., different spin or structural phases 
\cite{tsuchiya06,hsu09}. 

\begin{table}
\caption{\label{niostruct}
The equilibrium lattice parameter, ($a$, in Bohr atomic radii), 
the bulk modulus ($B$, in GPa), 
and the band gap ($E_g$, in eV) of NiO obtained with different 
computational approaches: GGA, ``traditional" GGA+U (with U only
on the $d$ states of Ni), GGA+U+V and a ``self-consistent" 
GGA+U+V with the interaction parameters computed from a GGA+U+V
ground state (see text). Comparison is made with the experimental 
results on all the computed quantities.}
\begin{ruledtabular}
\begin{tabular}{cccc}
 & $a$ & $B$ & $E_g$  \\
\hline
GGA & 7.93 & 188 & 0.6  \\
GGA+U & 8.069 & 181 & 3.2   \\
GGA+U+V & 8.031 & 189 & 3.6  \\
GGA+U+V$^{sc}$ & 7.99 & 197 & 3.2  \\
Exp & 7.89\footnotemark[1] & 166-208\footnotemark[2] & 3.1-4.3\footnotemark[3]  \\
\end{tabular}
\end{ruledtabular}
\footnotetext[1]{Reference \cite{crc98}}
\footnotetext[2]{Reference \cite{huang94}}
\footnotetext[3]{Reference \cite{fujimori84,sawatzky84}}
\end{table}

\begin{figure}
\vspace{-0.1in}
\includegraphics[width=5.5cm,keepaspectratio,angle=-90]{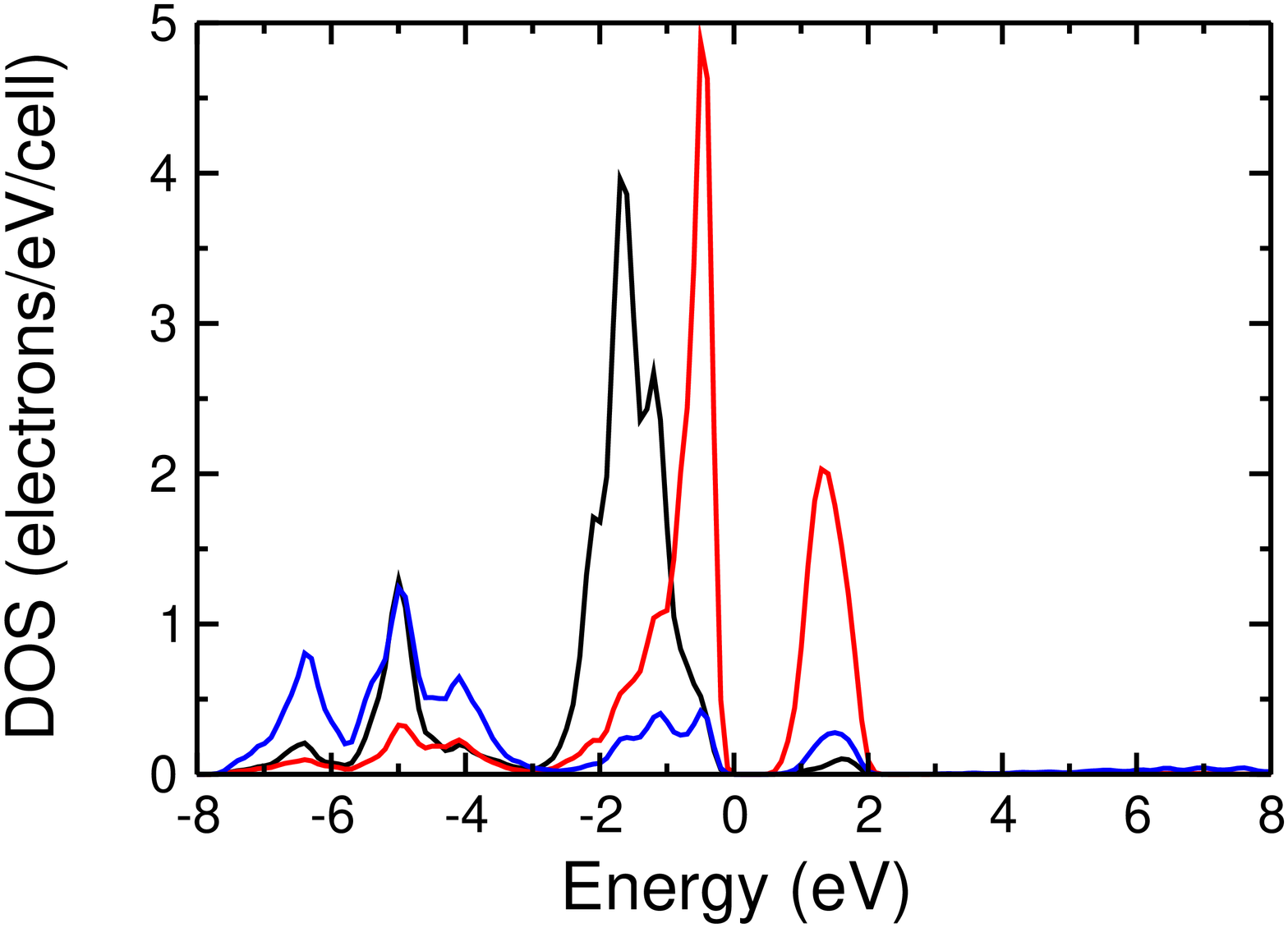}
\vspace{-0.1in}
\includegraphics[width=5.5cm,keepaspectratio,angle=-90]{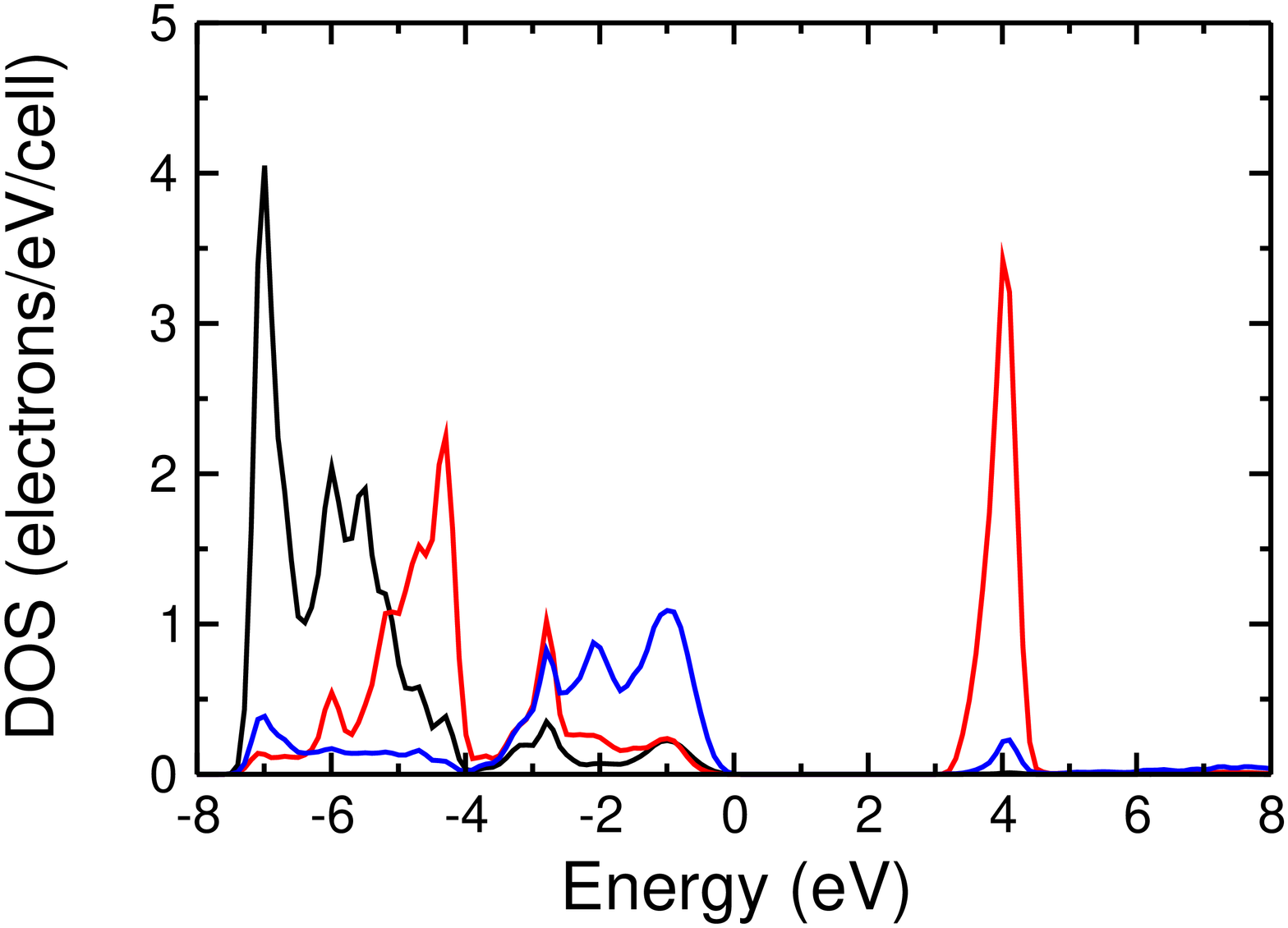}
\vspace{-0.1in}
\includegraphics[width=5.5cm,keepaspectratio,angle=-90]{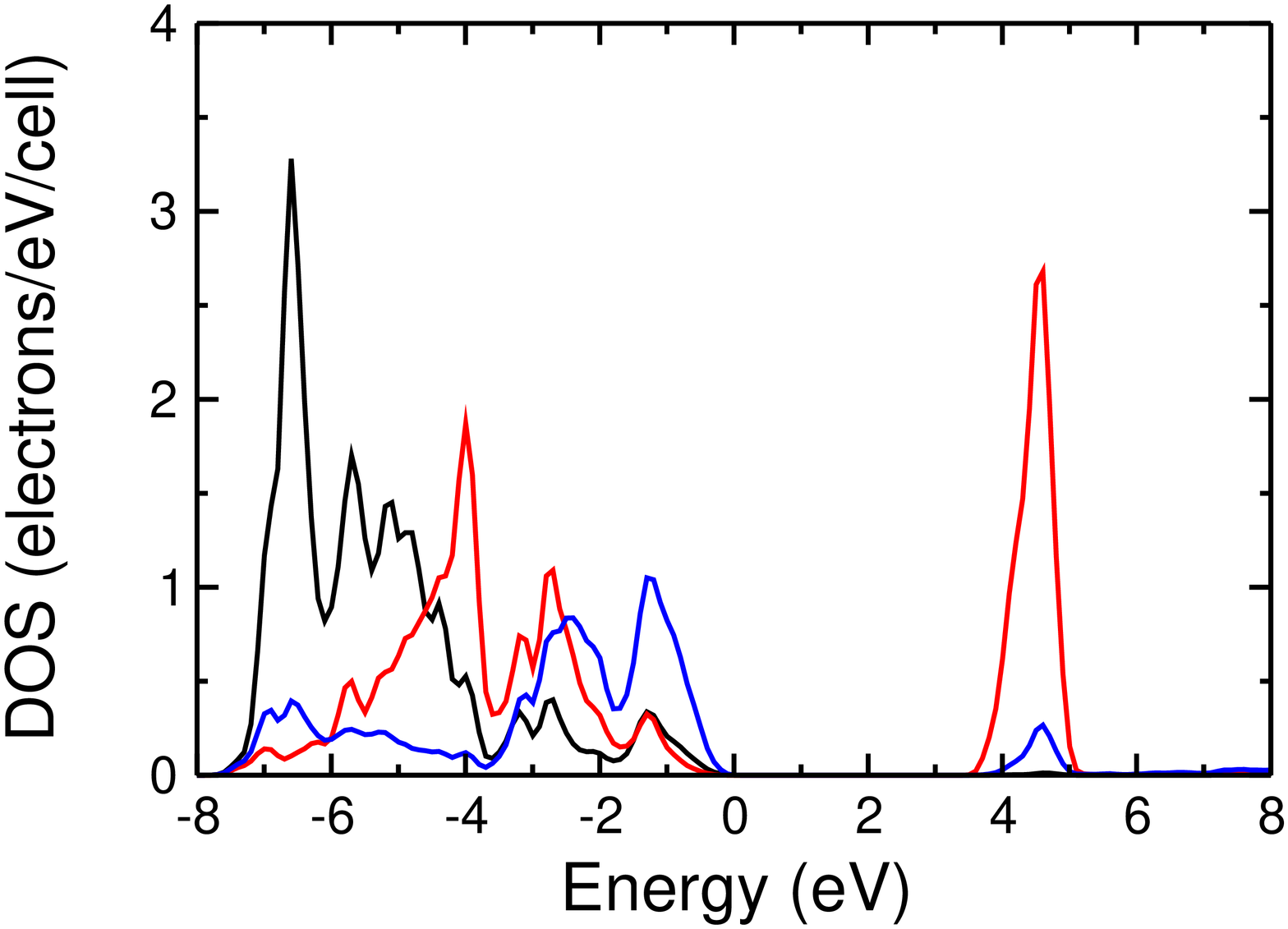}
\vspace{-0.1in}
\includegraphics[width=5.5cm,keepaspectratio,angle=-90]{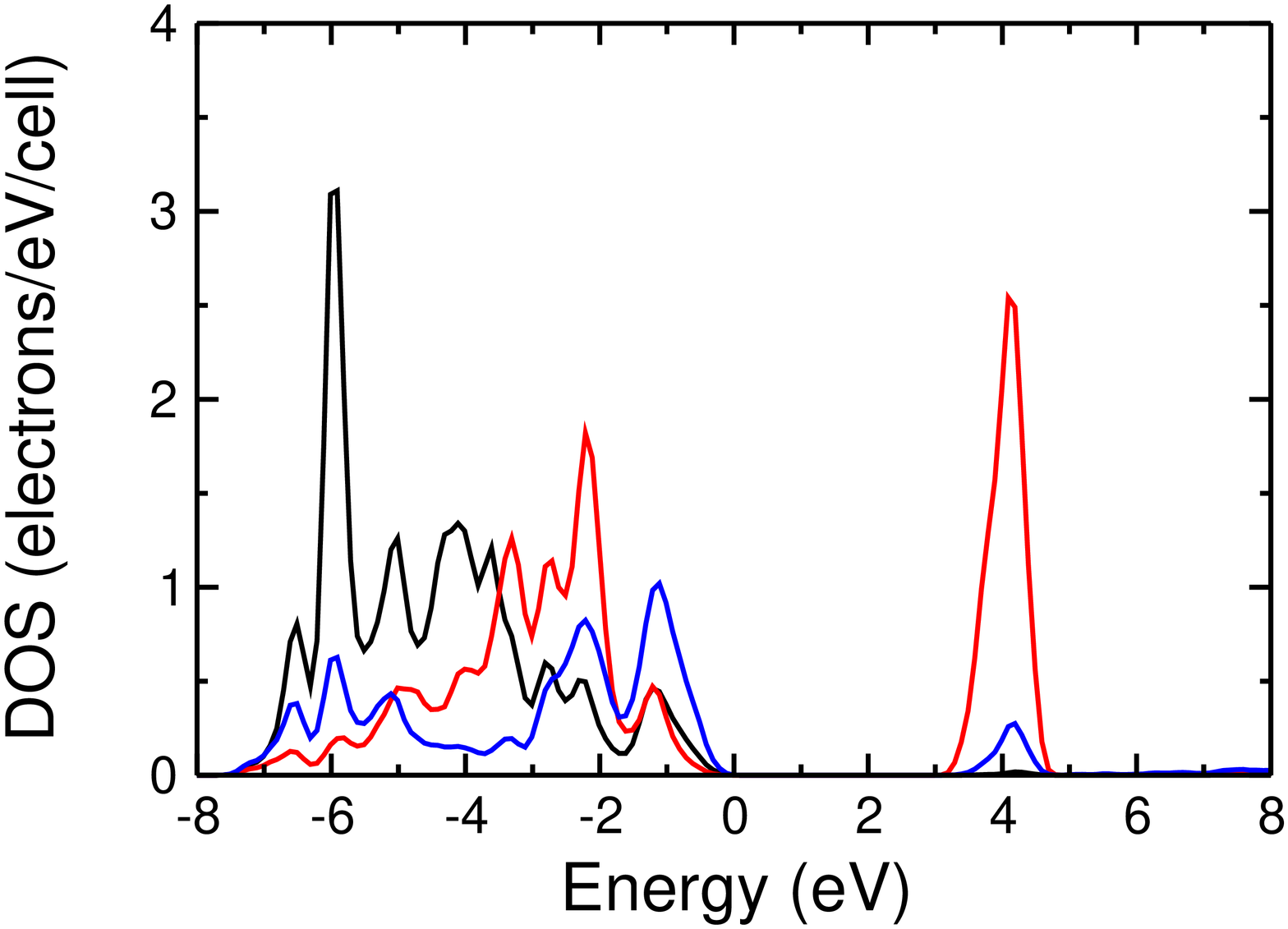}
\caption{\label{niodos} (Color online) The Density of states of NiO obtained 
with different approximations: a) GGA; b) GGA+U ($U$ on Ni $d$ states 
only); c) GGA+U including on-site interactions on Ni $d$ ($U_{Ni}$)
and O $p$ ($U_{O}$) states; d) GGA+U+V including on-site
($U_{Ni}$ and $U_{O}$) and inter-site ($V_{Ni-Ni}$, $V_{O-O}$,
and $V_{Ni-O}$) interactions. Only ``standard" states corrected.
In all graph the black and red lines represent majority and minority
spin $d$ states of Ni, the blue line the $p$ states of O.}  
\end{figure}

Table \ref{niostruct} compares the
equilibrium lattice parameters,  the bulk moduli, and the energy
band gap obtained in the
three different approaches with results from experiments. 
In each case the band gap of the material was measured for the equilibrium 
lattice parameter reported in the same table.
A comparison between the DOS obtained
within the different approaches is made in Fig \ref{niodos}.

Our GGA calculations
result in an equilibrium lattice spacing of 7.93 bohrs which is 
quite consistent (+0.5\%) with the experimental value of 7.89 bohrs and 
provides a good estimate of the bulk modulus (187.7 GPa).
As expected from other GGA results, 
the band gap ($\sim 0.6$ eV, Fig.~\ref{niodos}, top graph) 
is smaller than the observed one (3.0 - 4.3 eV)
and its spectroscopic nature is also not consistent with experiments
as the top of the valence band is dominated by the $d$ states of Ni instead
of O $p$.
It is important to notice that the peak at 5 eV below the edge of the
valence band contains contributions from both O and Ni and,
as specified in other works in literature (see, e.g.,
\cite{kressenio04}), is the signature of
the hybridization between $d_{e_{g}}$ and $p$ states.
This is a spurious feature, common also to DFT+U calculations,
as the lowest part of the valence state should be dominated
by $t_{2g}$ states instead \cite{sawatzky84,thuler83,
fujimori84}. 

GGA+U ($U$ on Ni $d$ states only) leads to a significant 
improvement in the size of the 
band gap (3.2 eV) compared to GGA 
and O $p$ states are dominant on the top
of the valence band in agreement with experimental results 
(Fig.~\ref{niodos} second graph from the top).
The equilibrium lattice parameter obtained with GGA+U (8.069 bohrs) 
is larger than the one obtained 
with GGA, deteriorating the agreement with experiments
(table \ref{niostruct}).
Also, the crystal results softer with a bulk modulus of about 181 GPa
(still in the range of experimental results).
While the almost exclusive presence of $p$ bands on top of the valence
band seems to overestimate the spectral weight of these states compared with
the results of other calculations 
\cite{dudarevnio00,bengone00,kressenio04}, the dominant $p$ character
of the peak at -2 eV, and the appearance of a peak at -4.5 eV, 
mainly due to minority-spin $t_{2g}$ Ni states, 
have not been highlighted in experiments \cite{fujimori84}.
However, the position of a strong peak in the DOS at about -7 eV from the 
valence band edge appears consistent with experimental results 
\cite{sawatzky84} although further analysis would be needed to 
assess its character.
The clear-cut separation between $p$ states
and lower energy $d$ states is, instead, unusual for 
DFT+U calculations.
This latter aspect may be due to the different value of the $U$ parameter
used with respect to other calculations of this kind.

Using GGA+U+V, the equilibrium lattice parameter (Table \ref{niostruct}) 
is corrected towards the experimental value (8.03 bohrs) 
while the bulk modulus results equal to
189 GPa and a band gap of about 3.6 eV opens between O $p$ and Ni $d$ states
correctly placed at the top of the valence band and the bottom of the
conduction one respectively.
The band structure of the system shows some differences from the one obtained
with GGA+U. 
With respect to the GGA+U results,
the minority spin $d$ states of Ni have moved
upwards in energy and show a larger overlap with the $p$ states of O
that suggests a larger degree of hybridization.
As a result, the peak at -2.5 eV acquires a markedly mixed
$d$ and $p$ character, and the 
central peak, mainly due to Ni minority-spin $d$ states, 
moves to slightly higher energy at -4 eV from the gap.
The occupied satellite peak, that also appears at slightly higher energy 
($\approx$ -6.5 eV) than in the GGA+U DOS, 
is still dominated by Ni majority-spin $d$ states.

Using the ``self-consistent" GGA+U+V (i.e., with $U$ and $V$ computed
from a GGA+U+V ground state) we obtained an equilibrium
structure in better agreement with experimental results ($a = 7.99$ bohrs),
a bulk modulus of about 197 GPa (still within the range of available
experimental data) and a band gap of 3.2 eV (at the basis of the valence 
and conduction peaks) as evident from
the bottom graph of Fig.~\ref{niodos}. 
The electronic structure in this case is quite similar to that
obtained within the non self-consistent GGA+U+V approach (Fig.~\ref{niodos},
third graph from top).
However an even larger overlap can be observed between Ni $d$ 
and O $p$ states in the upper part of the spectrum
with a peak below the edge of the valence
band at $\approx$ -2.2 eV that is dominated by Ni states
in agreement with experiments \cite{sawatzky84}.
The occupied satellite peak, at about -6 eV, is still dominated
by Ni $d$ states although the $p$ state component seems slightly 
more significant in this case.
It is important to notice that the valence DOS has now acquired
three dominant features that correspond to the O peak at the
top of the band, a Ni-dominated peak at about -2.2 eV and a strong
$d$ peak at -6 eV. These characteristics are consistent with the
results from photoemission experiments although the relative spectral 
weight of these peaks are not in quantitative agreement with observations.
Furthermore we don't observe the downshift of the occupied satellite
(actually moved to slightly higher energies)
that in Ref.~\cite{miura08} is expected to descend from the explicit 
inclusion of inter-site electronic interactions in the corrective functional.   
In our opinion these discrepancies are 
due to the fact that, at the present level of 
approximation, electronic interactions parameters are only site-dependent.
While dynamical correlation may play an important role, as demonstrated by
DFT+DMFT calculations \cite{vollhardt06,kunes07,miura08}, we believe
that the most significant refinement needed to reproduce all the
details of the spectrum of the material consists
in the formulation of an extended model 
with orbital-dependent interaction parameters.
Specifically, a distinction should be made between the (inter-site)
interactions of O $p$ states with Ni $t_{2g}$ and $e_g$ states.

In summary, while not completely satisfactory, GGA+U+V offers a 
quite significant improvement in the description of NiO compared
to GGA+U and we believe that the more realistic representation of 
the electronic properties (as well as structural ones) can be 
a better starting point than GGA+U for numerical approaches (as GW
or DFT+DMFT) aimed 
at predicting the excitation spectrum of the material or at studying the
effects of dynamical correlations.
 
\subsection{Si and GaAs}

For Si and GaAs,
DFT calculations based on LDA and GGA 
approximate functionals provide results for the structural properties 
(e.g., the lattice parameter and the bulk modulus and its first derivative)
\cite{nielsen85,filippi94,juan95}
and the vibrational spectrum \cite{giannozzi91} in 
good to excellent agreement with experiments.
GGA normally results
in slightly longer equilibrium lattice constants and softer bulk
moduli than the measured ones \cite{filippi94,juan95,lee97}. 
However, the computed band gap of these materials 
systematically underestimates the experimental value
due to inaccuracies inherent to approximate DFT energy functionals
and, in particular, to the lack of finite discontinuities
in the exchange-correlation potential \cite{perdew83,grosslibro}.
Many corrective approaches have been developed to overcome these difficulties
of standard DFT approximations and quite good agreement with the
experimental spectrum could be obtained from numerical approaches
correcting the exchange correlation energy model, as in exact-exchange 
(EXX) \cite{stadele99,sdg09} or hybrid functional \cite{heyd05} calculations, 
and from methods specifically designed to reproduce the electronic excitation
spectrum, as the GW approach based on an LDA \cite{rohlfing93,aulbur99} or a 
EXX \cite{aulbur00} ground state.
In the present work, we investigate the performance 
of the GGA+U+V functional on Si 
and GaAs and compare its results with those from standard GGA and GGA+U. 
In these semiconductors, the hybridization between $s$ and $p$ 
orbitals and the formation of covalent bonds between neighbor atoms are 
key ingredients to understand their structural and electronic properties. 
Thus, the corrective functional introduced in the present paper,
including inter-site interactions and acting on two angular momenta
manifolds per atom, contains all the necessary ingredients 
to achieve significant improvement with respect to standard DFT 
approximations in the description of these materials.

As mentioned above, for both Si and GaAs we have used GGA 
exchange-correlation functionals constructed with the PBE parametrization
\cite{pbe}.  
In particular for Si we employed a norm-conserving pseudopotential
(Si.pbe-rrkj.UPF of the QE repository) that was built
on the $3s^2 3p^2 3d^0$ valence configuration.
This potential required an energy cut-off of 40 Ry for the 
plane-wave expansion of the Kohn-Sham wavefunctions. 
For GaAs we used, instead, 
ultrasoft pseudopotentials \cite{vanderbilt90}
that required cut-offs of 40 and 400 Ry for
the expansion of the electronic wavefunctions and charge density 
respectively.
For As we chose a pseudopotential constructed on the $4s^24p^3$ configuration
(As.pbe-n-van.UPF of the QE library). 
As reference \cite{lilienfield08} points out, predictions for gallium compounds
can be very sensitive to the inclusion of $3d$ electrons in the
valence manifold for Ga pseudopotentials. 
Thus, two alternatives were 
considered for Ga with $3d$ electrons in valence ($3d^{10}4s^24p^1$ - 
Ga.pbe-nsp-van.UPF from the QE repository), 
or frozen in the core ($4s^24p^1$). 
We will refer to these two different situations as GaAs(v) and
GaAs(c) respectively. The Ga pseudopotential for GaAs(c)
was constructed using D. Vanderbilt's code and library \cite{vanderbiltcode}.
For both materials (Si and GaAs), the Brillouin zone was sampled
with a $8\times8\times8$ Monkhorst-Pack \cite{monkhorst76} k-point grid.

\begin{figure}
\includegraphics[width=6.5cm,keepaspectratio,angle=-90]{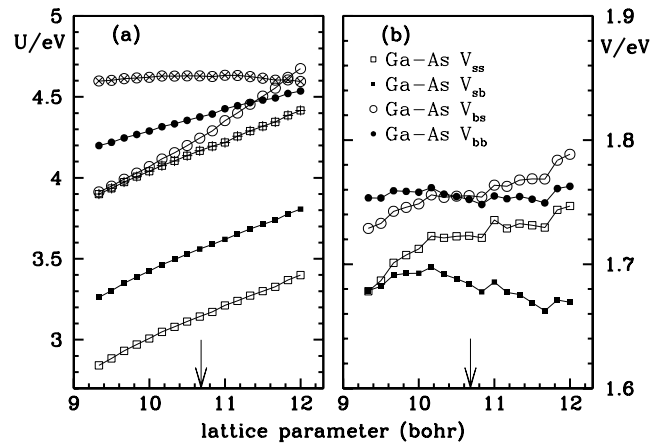}
\caption{ Intra-site ($U$) and inter-site ($V$) interaction
parameters as functions of lattice parameter ($a$) for GaAs with $3d$ 
electrons as valence electrons. a) Intra-site interactions. Squares 
correspond to Ga atom and circles to As atom. Filled symbols represent 
$U_{ss}$ (standard-standard), open symbols represent $U_{sb}$ 
(standard-background) and symbols with crosses 
represent $U_{bb}$ (background-background). b) Inter-site interactions. 
Arrows indicate the equilibrium lattice parameter.}
\label{fig1}
\end{figure}

\begin{table}
\caption{\label{table_UV}Interaction parameters $U$ and $V$ (eV) for Si and
GaAs\protect\footnotemark. 
Inter-site terms are for first-neighbors and the 
listed values are for the equilibrium lattice parameters found with GGA+U+V. 
Indexes $s$ and $b$ stand for ``standard'' and ``background'' 
orbitals respectively as discussed in the text.}
\protect\footnotetext{Ga $3d$ electrons as valence electrons.}
\begin{ruledtabular}
\begin{tabular}{ccccccccc}
 & $U_{ss}$ & $U_{sb}$ & $U_{bs}$ & $U_{bb}$ & $V_{ss}$ & $V_{sb}$ & $V_{bs}$ &
$V_{bb}$ \\
\hline
Si-Si & 2.82 & 3.18 & 3.18 & 3.65 & 1.34 & 1.36 & 1.36 & 1.40 \\
Ga-Ga & 3.14 & 3.56 & 3.56 & 4.17 &      &      &      &      \\
As-As & 4.24 & 4.38 & 4.38 & 4.63 &      &      &      &      \\
Ga-As &      &      &      &      & 1.72 & 1.68 & 1.76 & 1.75 \\
\end{tabular}
\end{ruledtabular}
\end{table}

In order to determine the equilibrium lattice parameter and the bulk modulus,
the dependence of the ground-state energy on volume was fitted by the
Murnaghan equation of state. 
As in the case of NiO, DFT+U and DFT+U+V total energy calculations were
performed using interaction parameters consistently recomputed
for each value of the lattice parameter.
Fig.~\ref{fig1} illustrates the dependence of $U$s and $V$s on the size 
of the unit cell in the case of GaAs(v).

In Table~\ref{table_UV}, the on-site and inter-site interactions computed 
at the equilibrium lattice parameter are grouped to ease the comparison. 
It is important to notice that the values of either the $U$s or the $V$s are 
very similar irrespective of the orbital manifold the 
parameters acts on ($p$ or $s$ states). 
This result, more evident for GaAs and for the inter-site couplings,
is due to the hybridization of $p$ and $s$ orbitals and corroborates 
the need of treating both orbitals at the same level.

\begin{table}
\caption{\label{table_comp}Comparative results for lattice 
parameter ($a$, in ${\rm \AA}$), bulk modulus (B, in GPa) 
and energy gap ($E_g$, in eV).}
\begin{ruledtabular}
\begin{tabular}{cccc}
& Si & GaAs(v)\footnotemark[1]  & GaAs(c)\footnotemark[2] \\
\hline
& $a$\hspace{3.6mm} B\hspace{3.6mm} $E_g$ & $a$\hspace{3.6mm} B\hspace{3.6mm} $E_g$ & $a$\hspace{3.6mm} B\hspace{3.6mm} $E_g$ \\
\hline
GGA & 5.479,~83.0,~0.64 & 5.774,~58.4,~0.19 & 5.578,~65.7,~1.25\\
+U & 5.363,~93.9,~0.39 & 5.736,~52.6,~0.00  & 5.616,~62.7,~0.81\\
+U+V & 5.370,~102.5,~1.36& 5.654,~67.7,~0.90 & 5.535,~76.5,~1.97\\
Exp.\footnotemark[3]& 5.431,~98.0,~1.12 & 5.653,~75.3,~1.42  & 5.653,~75.3,~1.42\\
\end{tabular}
\end{ruledtabular}
\footnotetext[1]{$Ga$ $3d$ electrons as valence electrons.}
\footnotetext[2]{$Ga$ $3d$ electrons in the core.}
\footnotetext[3]{at $300$~K, from Ref.~\cite{semicondbase}. At $0$~K, 
the energy gaps of Si and GaAs are estimated to be $1.17$~eV and $1.52$~eV 
respectively~\cite{semicondbase}.}
\end{table}

In Table \ref{table_comp}, the equilibrium lattice parameter, the 
bulk modulus and the band energy gap obtained from GGA, GGA+U and GGA+U+V
calculations on Si and GaAs can be directly compared with experimental 
measurements of the same quantities (we refer to the data collected
in the web-database, Ref. \cite{semicondbase}). 
As expected from abundant literature, GGA overestimates the
equilibrium lattice parameter with respect to the
experimental value (except for the case of GaAs(c)),  
while the bulk modulus and the band gap 
are underestimated in all cases. 
On-site GGA+U predicts the equilibrium lattice 
parameter in better agreement with the
experimental value (overcorrected for Si); 
however, the bulk modulus is improved with respect to
the GGA value only in the case of Si. 
In all three cases, however, the energy band gap is lowered compared to GGA,
further worsening the agreement with experiments.
While this result may appear strange, it is not unexpected. In fact 
the on-site corrective functional suppresses the hybridization between
states on neighbor atoms that is largely responsible for the 
band gap in semiconductors between valence 
(bonding) and conduction (anti-bonding) states.
The use of the inter-site correction, in spite of the fact that 
$V$s are smaller than half the on-site $U$s, results in a
systematic improvement for the evaluation of all these quantities.
In fact, encouraging the occupations of hybrid states, the inter-site
interactions not only enlarge 
the splittings between populated and empty orbitals (which increases
the size of the band gap), but also make bonds shorter (so that 
hybridization is enhanced) and stronger, thus tuning both the equilibrium
lattice parameter and the bulk modulus of these materials to 
values closer to experimental results.
For GaAs the calculations with the Ga $3d$ states in valence are 
more accurate than the ones with these atomic states frozen in the
core. In fact, for GaAs(v) we obtain an equilibrium lattice parameter
in excellent agreement with the measured one, while bulk modulus and 
band gap slightly underestimate the experimental results, yet improving
the estimates obtained with GGA and GGA+U. For GaAs(c), instead,
while the bulk modulus is in very good agreement with experiments,
the equilibrium lattice spacing is underestimated and the band gap
significantly overestimated worsening, for both quantities, the 
accuracy of the results obtained with GGA and GGA+U. Our results thus
confirm that Ga $3d$ should be treated as valence states; 
in the present study, however, they are not directly subject to the
Hubbard functional.

Despite the overall improvement obtained with GGA+U+V, some 
discrepancies with the experimental results still 
persists regarding, especially, the equilibrium lattice 
constant and the bulk modulus. 
However, it should be 
kept in mind that computational results presented in this
section would be directly comparable
to 0 K measurements for which slightly shorter lattice parameters 
and slightly higher bulk moduli are predicted \cite{semicondbase}
(see footnotes of Table \ref{table_comp} for details).

\begin{figure}
\includegraphics[width=5.5cm,keepaspectratio,angle=-90]{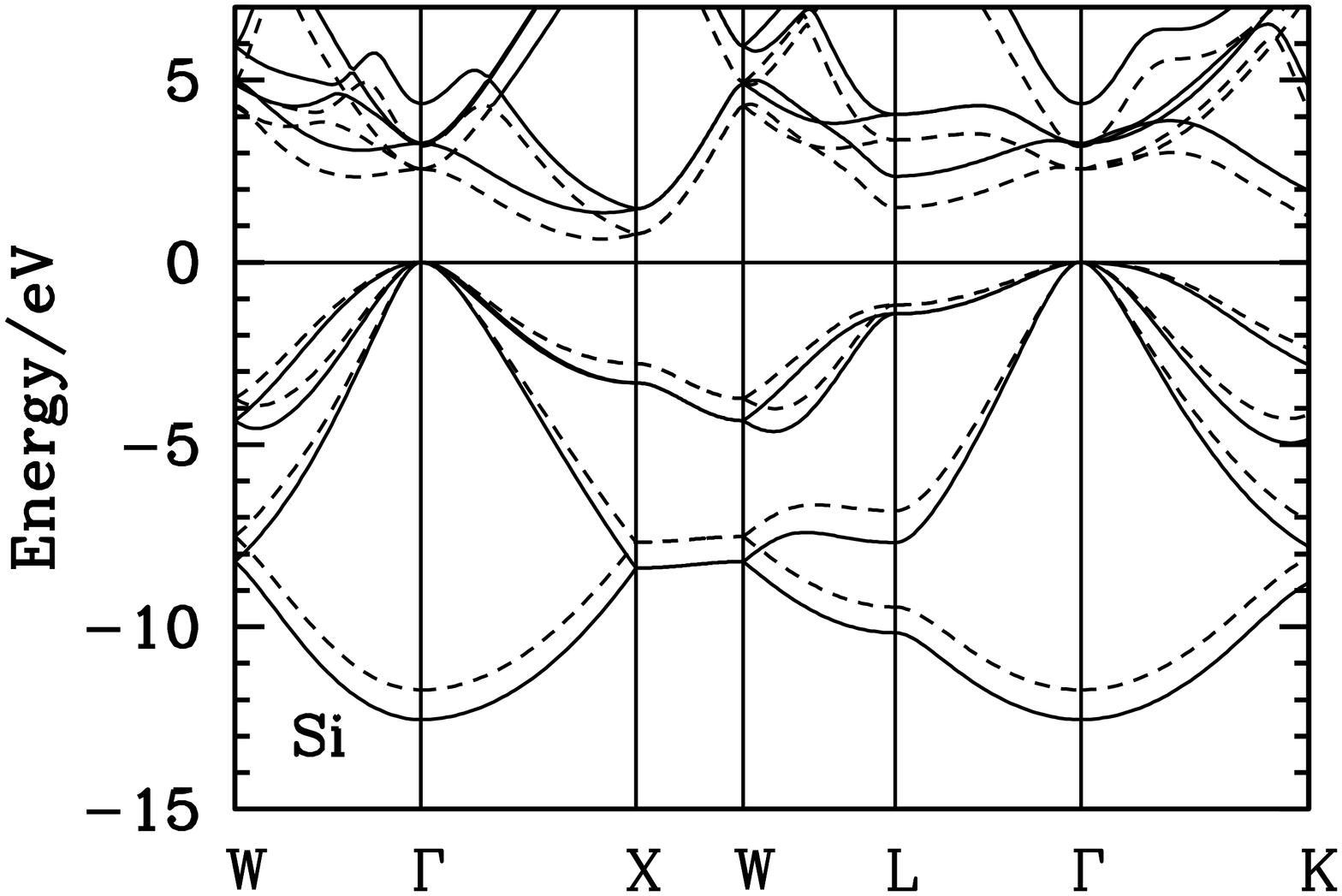}%
\vspace{-0.1in}
\includegraphics[width=5.5cm,keepaspectratio,angle=-90]{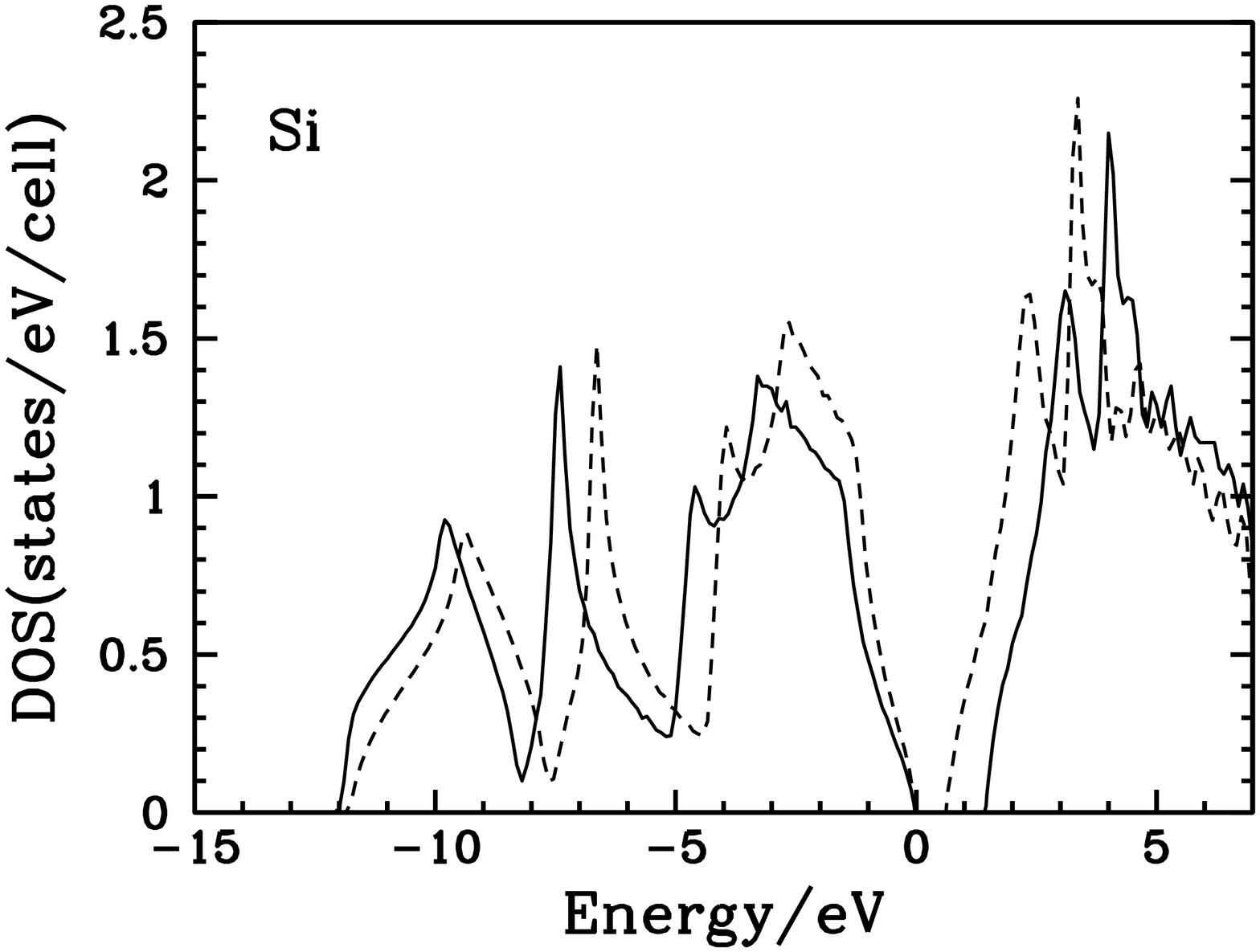}%
\vspace{-0.1in}
\includegraphics[width=5.5cm,keepaspectratio,angle=-90]{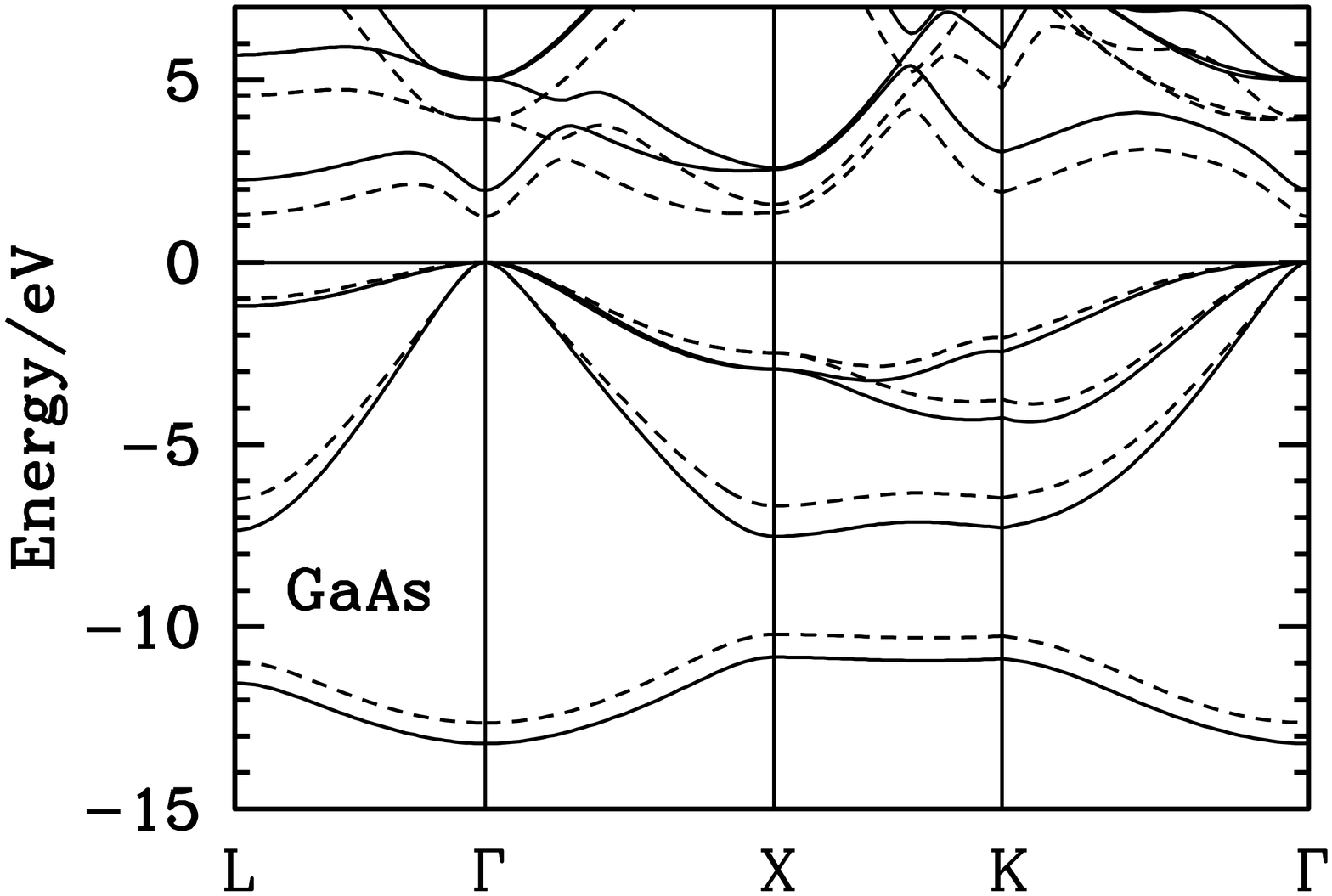}%
\vspace{-0.1in}
\includegraphics[width=5.5cm,keepaspectratio,angle=-90]{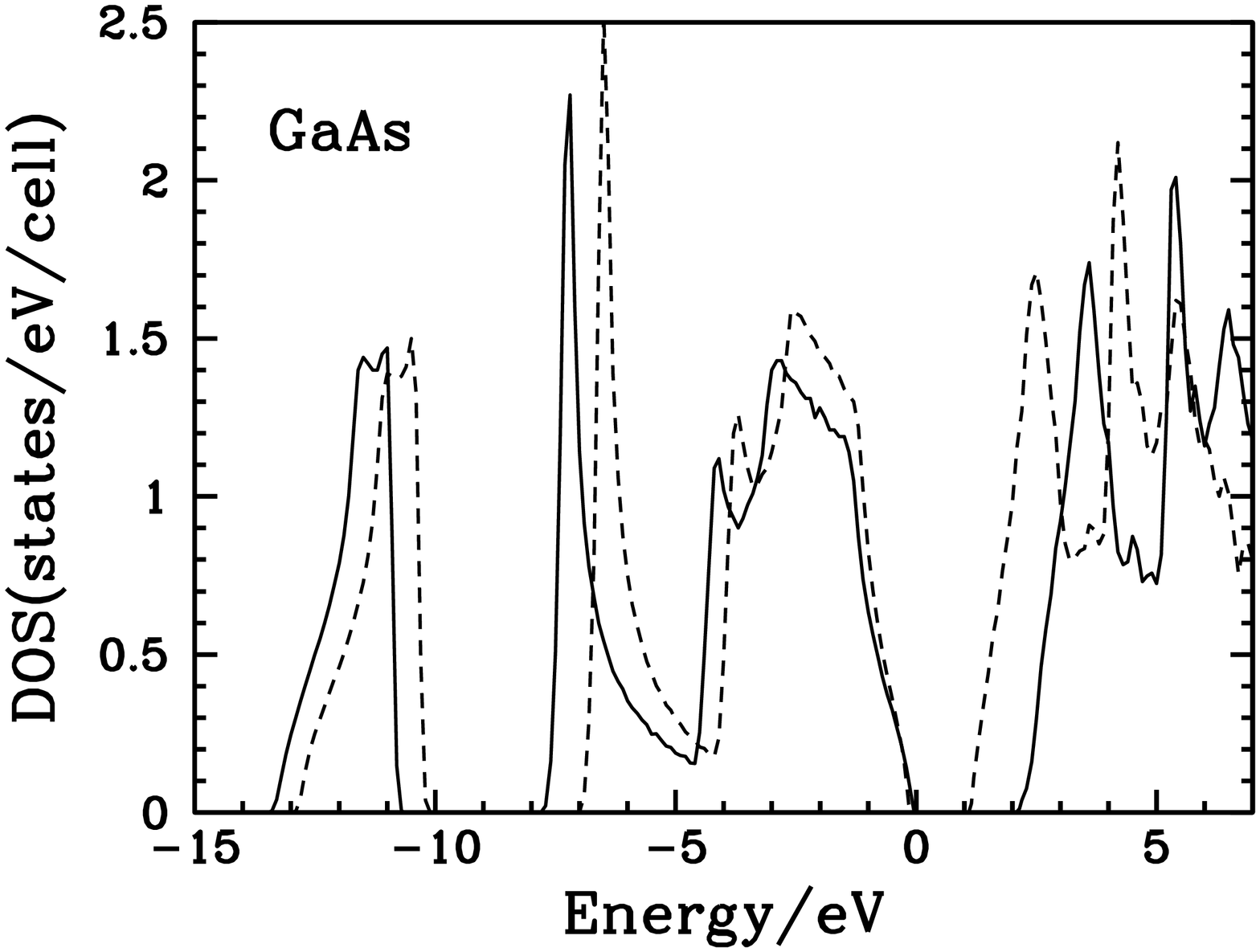}%
\caption{Band structure and density of states of Si and GaAs (core case). Continuous lines represent GGA+U+V results and dashed lines represent standard GGA results. All energies were shifted so that the top of valence bands are at zero energy.}
\label{fig2}
\end{figure}

Fig.~\ref{fig2} displays the GGA and GGA+U+V band structures and 
density of states for Si and GaAs. The energies were shifted so that
the top of the valence band corresponds to the zero of the energy 
in all cases.
The correction introduced by the GGA+U+V 
(with respect the GGA band structure) to the conduction band 
consists in an almost rigid upward shift in energy. 
In the valence manifold, instead, the correction
to the energy levels acquires a slight $k-$dependence and has a more
pronounced effect on the lowest energy level, which is shifted downward.
This latter effect results in an increase of the total bandwidth of these
systems (by 0.6 eV for GaAs and by 0.8 eV for Si) compared to GGA, and 
accounts for perfect agreement with experimental results (measured bandwidth is 12.5 eV for Si and 13.2 eV
for GaAs \cite{rohlfing93}).

In summary, the GGA+U+V shows significant improvement over GGA and GGA+U
in the description of structural and electronic properties of
Si and GaAs. 
Although the dominant effect arises from the interaction
between nearest neighbors sites, for both Si and GaAs the inclusion
of the coupling between further neighbors could bring
a refinement of the results presented in this section. 
Also, in the case of GaAs with Ga $3d$ electrons in valence,
more accurate results could be obtained accounting for the hybridization
of these states with $4s$ and $4p$ orbitals 
(probably the overlap with the states of As is quite limited) and 
including the corresponding interaction parameter
in the ``+U+V" corrective functional.
Since the GGA and the GGA+U+V ground states are qualitatively similar,
we expect a ``self-consistent" evaluation of $U$ and $V$
to have a minor effect on the presented results 
both for Si, and GaAs.

\section{Conclusions}

In this paper we have introduced a useful generalization (named DFT+U+V) 
of the popular DFT+U
method that, modeled on the extended Hubbard Hamiltonian, 
includes on-site ($U$) and inter-site
($V$) electronic interactions and is able to correct up to two angular momentum
manifolds per atom. 
The competition between the two kinds of electronic couplings 
avoids the over-stabilization of occupied atomic
orbitals, often affecting the on-site DFT+U, and
allows for the description of more general ground states where 
electrons ``localize" on hybridized (e.g., molecular) 
orbitals.
The flexibility in the representation of occupied states is further enhanced
by the larger number of orbital manifolds that are simultaneously
subject to the action of the corrective functional.
Numerical accuracy is guaranteed by the linear-response 
calculation of $U$ and $V$ that allows to 
evaluate the relative strength of the two couplings and, 
thus, to precisely determine the degree of electronic localization 
of the ground state.

The effectiveness of the method is demonstrated in this paper
by its successful application to 
quite diverse test systems as NiO (Mott/charge-transfer
insulator), Si and GaAs (band semiconductors).
For both classes of materials the use of the DFT+U+V functional results in 
a significant improvement (over approximate DFT and DFT+U) in the agreement 
of structural (equilibrium lattice 
parameter and bulk modulus) and electronic (overlap of 
$p$ and $d$ states, band gap) properties with available 
experimental results. For Si and GaAs this improvement is 
more consistent, probably due to the higher level of structural
isotropy of these materials that fits better with 
a corrective functional with orbital independent interactions.

The importance of these results is not limited to the more accurate 
description the selected test systems received. 
In fact, they demonstrate that the novel 
computational approach provides a unified theoretical framework able
to treat systems as divers as Mott, charge-transfer, and band insulators
with equal accuracy and comparable computational effort. 
Furthermore, the use of the extended Hamiltonian provides a new route to
improve the corrective functional in a systematic way as it allows to add
interactions between further shells of neighbors independently and, thus,
to easily assess their individual contributions to the final results.
This procedure also leads to the possibility to set calculations of
minimal cost (with a minimal set of inter-site interaction terms).

Potential applications of this novel corrective functionals are quite 
broad. Besides high T$_c$ superconductors (for which 
the nearest-neighbor electronic couplings has sometimes been demonstrated
to play a very important role), the remarkable success
obtained with strongly localized (correlated) and strongly hybridized
systems, suggests that intermediate situations where ``correlated" and 
``non-correlated" orbitals show significant overlap (as, e.g., 
semiconductors doped with magnetic impurities, metallic 
active centers of molecular complexes, etc) and 
phenomena where a switch between different localization regimes is observed
(e.g., in bond breaking and formation events) are likely to be described
quite accurately by the novel DFT+U+V scheme. 
This is especially true for isotropic systems that better fit
the simple ``+U+V" functional based on orbital-independent interactions; 
more anisotropic materials may require, instead, extra attention. 
The development of a more flexible corrective Hamiltonian
based on orbital-dependent effective interactions, needed in these cases, is
planned for future investigations.

\section{Acknowledgments}

M. C. thanks Drs P. M. Grant, A. Floris, E. K. U. Gross and S. de Gironcoli, 
for very useful discussions. 
The authors are also grateful to the Minnesota Supercomputing Institute
for providing computational resources and technical support that 
were essential for this work.

\bibliography{paper_Campo_Cococcioni}

\end{document}